\documentclass{ecai}

\usepackage{graphicx}
\usepackage{latexsym}
\usepackage{multirow}
\usepackage{graphicx}
\usepackage{subfigure}
\usepackage{amsmath}
\usepackage{amssymb}

\begin{document}

\begin{frontmatter}

\title{Dual-Scale Interest Extraction Framework with Self-Supervision for Sequential Recommendation}
\author[A]{\fnms{Liangliang}~\snm{Chen$^{\dagger}$}}
\author[B]{\fnms{Hongzhan}~\snm{Lin$^{\dagger}$}\orcid{0000-0002-4111-8334}}
\author[A]{\fnms{Jinshan}~\snm{Ma}}
\author[A]{\fnms{Guang}~\snm{Chen}\thanks{Corresponding Author. Email: chenguang@bupt.edu.cn. This work was partially supported by MoE-CMCC "Artificial Intelligence" Project No. MCM20190701. ${\dagger}$ The first two authors contributed equally to this work.}}

\address[A]{Beijing University of Posts and Telecommunications}
\address[B]{Hong Kong Baptist University}

\begin{abstract}
 In the sequential recommendation task, the recommender generally learns multiple embeddings from a user's historical behaviors, to catch the diverse interests of the user. Nevertheless, the existing approaches just extract each interest independently for the corresponding sub-sequence while ignoring the global correlation of the entire interaction sequence, which may fail to capture the user's inherent preference for the potential interests generalization and unavoidably make the recommended items homogeneous with the historical behaviors. In this paper, we propose a novel Dual-Scale Interest Extraction framework (DSIE) to precisely estimate the user's current interests. Specifically, DSIE explicitly models the user's inherent preference with contrastive learning by attending over his/her entire interaction sequence at the global scale and catches the user's diverse interests in a fine granularity at the local scale. Moreover, we develop a novel interest aggregation module to integrate the multi-interests according to the inherent preference to generate the user's current interests for the next-item prediction. Experiments conducted on three real-world benchmark datasets demonstrate that DSIE outperforms the state-of-the-art models in terms of recommendation preciseness and novelty.
\end{abstract}

\end{frontmatter}

\section{Introduction}

\begin{figure}[t]
  \centering
  \scalebox{1.0}{\includegraphics[width=\linewidth]{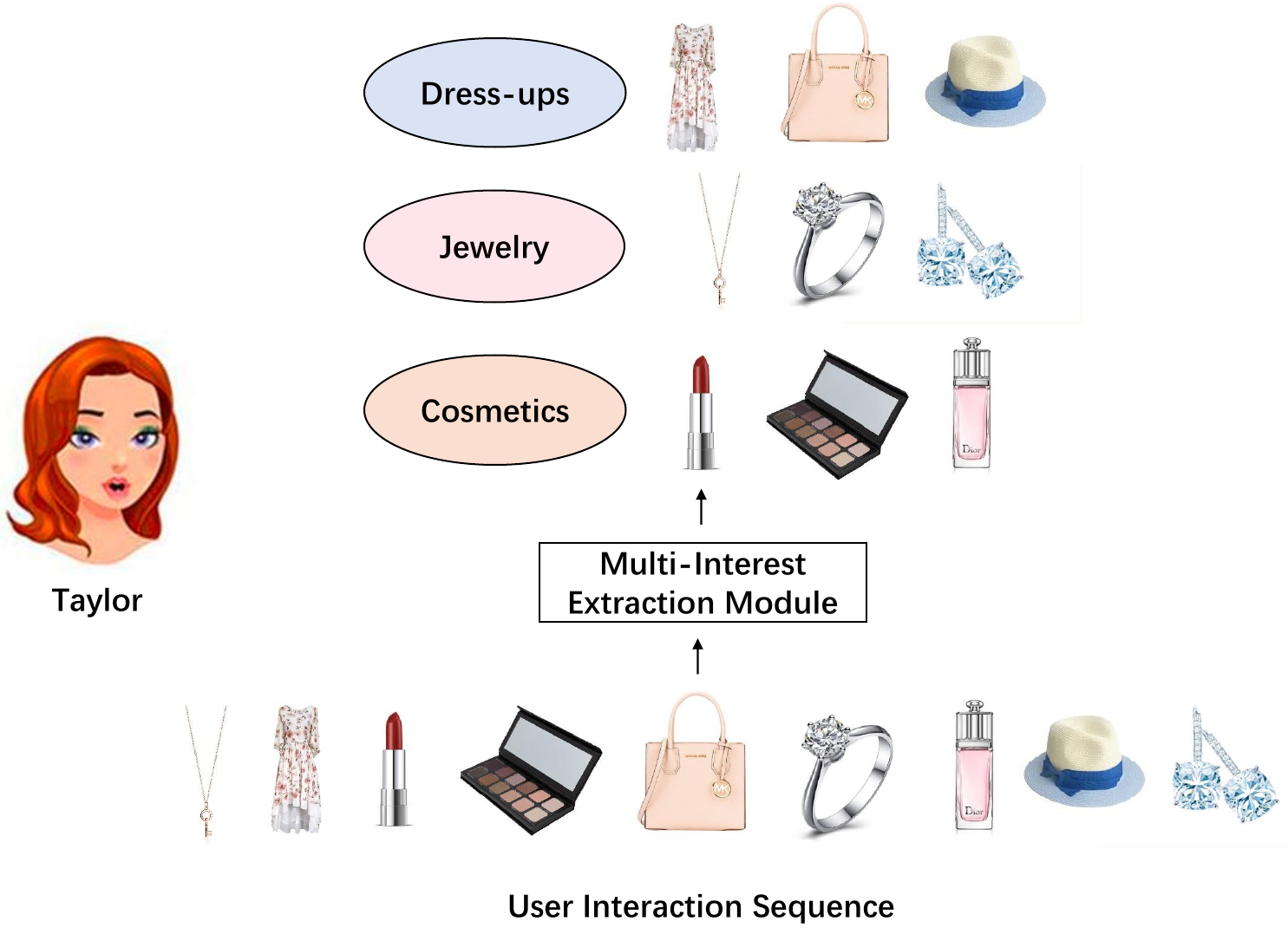}}
  \caption{An illustration of a user's diverse interests in the e-commerce platform, Taylor has multiple interests including dress-ups, jewelry, and cosmetics. The recent multi-interests extraction driven methods usually extract each interest from the sub-sequence (i.e., local scale) of the user's interaction sequence.} \label{multi-interest}
\end{figure}

Sequential recommender systems aim to characterize the user's current interests and predict the successive item based on his/her historical behaviors. For example, one might naturally purchase eggs and flour soon after consuming an oven. 
Equipped with the neural network's expressive power, the standard paradigm~\cite{hidasi2015session, tan2021sparse, cen2020controllable, kang2018self} for the sequential recommenders encodes his/her historical interactions to model the user's current interests for the recommendation.
Recent solutions \cite{li2019multi,cen2020controllable,tan2021sparse} believe that users may have very diverse interests and interact with items that belong to different item categories in practice. 
For example, as Fig.\ref{multi-interest} shows, Taylor is interested in dress-ups, jewelry, cosmetics, etc. Hence, the existing methods employ multiple embeddings to learn different aspects of the user's interests for state-of-the-art performance.

The existing multi-interest driven recommenders focus on each particular type of interest independently extracted from part items of a given interaction sequence, by leveraging capsule networks \cite{li2019multi,cen2020controllable}, intention prototype clustering \cite{ma2019learning}, etc. Despite the significant progress of these works, they however largely neglect or oversimplify the global correlation among the user's entire interacted items, which lack the awareness of the user's inherent preferences and may fail to jump out of the user's 
recent interacted item categories to explore potential user interests.
Intuitively, recommendation results that only contain homogeneous items with his/her historical interactions are not satisfying~\cite{zhu2018learning}, thus it is essential to increase the recommendation novelty.
As exemplified in Fig.\ref{multi-interest}, Taylor's interests in dressing up, jewelry, and cosmetics have consistently expressed "love beauty" as her inherent preference, and analogously she may be interested in nice phone cases that are diverse from her historical interaction items.
Such potential interests derived from the user's inherent preference are not fully explored by recent work due to its limited generalization from the independent feature learning.
Therefore, it's necessary to fully explore such potential interests derived from the user's inherent preference to facilitate the recommendation novelty of the sequential recommenders. 



In this paper, we hypothesize that the user's inherent preference based on the global correlation greatly influences the user's interests in the specific items. In light of this, we propose to effectively estimate the user's current interest via both the local and global scale interest extraction for the sequential recommendation. Besides the local scale interest extraction capturing the user's diverse interest in a fine granularity from the corresponding sub-sequences like the recent approaches, a global scale interest extraction is further needed to understand the user's inherent preference based on his entire interaction sequence, which helps to estimate the user's current interest precisely and explore his/her potential interests. 


To this end, we propose a novel $\mathbf{D}$ual-$\mathbf{S}$cale $\mathbf{I}$nterest $\mathbf{E}$xtraction framework named DSIE, to effectively estimate the user's current interest for the sequential recommendation. 
At the global scale, we adopt a multi-layer residual network (i.e., ResNet) to encode the user's entire interaction sequence as the user's inherent preference representation. 
Since the public recommendation data only contains recommendation results rather than the causes, whether inherent preference dominates a certain user-item interaction lacks explicit supervision. 
To tackle this challenge, we design a self-supervised contrastive learning task and augment samples via item shuffling. Assuming that the inherent preferences extracted from them are close, the pairwise contrastive task is designed to let augmented data from the same example be discriminated against others. 
At the local scale, we exploit a multi-interest extractor module to learn representations of multiple interests in fine granularity from the corresponding sub-sequences discovered via intention prototype clustering. Encouragingly, the noisy behaviors (e.g., sales promotions, exposure bias \cite{zheng2021disentangling}, and position bias \cite{jagerman2019model}) that are inconsistent with user's real interests will be filtered out when clustering. 
We further develop an interest aggregation module, which leverages the inherent preference to guide the multi-interests aggregation to generate the user's current interest. 
In this way, 
the inherent preference is injected into the fine-grained multi-interests to estimate the user's current interest, leading to improving the recommendation preciseness and novelty.
Experiments on three real-world datasets demonstrate that our DSIE outperforms the state-of-the-art baseline models by a substantial margin and achieves significant improvements in recommending novel items. 


To summarize, the contributions of our proposed framework are as follows:

\begin{itemize}
\item To our best knowledge, we are the first to estimate the user's current interest from a dual-scale paradigm that integrates multi-interest extraction and inherent preference modeling for the sequential recommendation.

\item  We propose to use the self-supervised learning to overcome the challenge of lacking explicit supervision when capturing the user's inherent preference.

\item  We develop an adaptive interest aggregation to integrate the user's multi-interests according to his/her inherent preference to retrieve the relevant items in the top-${\rm N}$ recommendation scenario.


\end{itemize}

\section{Related Work}

In this section, we will briefly review several lines of works closely related to ours, including sequential recommendation and multi-interest Recommendation.

\subsection{Sequential Recommendation}
Compared to general recommenders via uniform modeling of user-item relationships \cite{rendle2012bpr}, sequential recommenders try to make use of the exact order of the interaction sequence. Based on the assumptions that: the more recent interactions are more valuable. Many recent works about recommender systems focus on this problem. One line of works estimates the transition probability between items based on Markov Chains. For example, \cite{rendle2010factorizing} subsumed both a common Markov chain and the normal matrix factorization model for sequential basket data. \cite{wang2015learning} extends the FPMC model and employs a two-layer structure to construct a hybrid representation of users and items from the last transaction. Another line of works incorporates neural networks (e.g., RNNs, CNNs, and self-attention) into sequential recommenders. For instance, \cite{zhou2019deep} adopted two layers of GRU to model the evolution of user interests. \cite{tang2018personalized} treated the user interaction sequence as an "image" and applied CNNs to extract user interests. Recently, as self-attention has been widely used in language understanding tasks \cite{devlin2018bert}, some researchers have sought to use self-attention as the backbone of sequential modeling. \cite{kang2018self} proposed a unidirectional model SASRec using a casual attention mask. \cite{sun2019bert4rec} encoded the sequence bidirectionally and outperformed the 
unidirectional recommenders with the help of "masked language model" \cite{devlin2018bert}. Besides, there are some other works proposed to extract knowledge with an external memory component. 
For instance, \cite{lian2021multi} utilized a memory network for user interest modeling.

\subsection{Multi-Interest Recommendation}
Observing that a single user representation is insufficient to reflect user's multiple interests, some researchers proposed to adopt multiple embeddings to address this issue. The existing approaches can be summarized into two categories. The first type of methods resorts to powerful neural networks to implicitly extract the user's multiple intentions. For example, MIND \cite{li2019multi} proposed a Behavior-to-Interest (B2I) dynamic routing based on Capsule network \cite{sabour2017dynamic} for adaptively aggregating user's behaviors into interest representation vectors. 
MIMN \cite{pi2019practice} used memory networks to capture user interests from long sequential behavior data. The other type of methods relies on a set of intention prototypes~\cite{lin2021boosting, lin2023zero} identified via clustering to explicitly capture a user's multiple intentions. DisenRec \cite{ma2019disentangled} utilized latent prototypes to help learn disentangled representations for the recommendation. ComiRec \cite{cen2020controllable} is a recent representative work for extracting multiple interests, including two variants, ComiRec-DR and ComiRec-SA. ComiRec-DR is an improved version of MIND, updating the original dynamic routing method provided by CapsNet. ComiRec-SA applies the soft-attention mechanism for multi-interest extraction. SINE \cite{tan2021sparse} maintained a large-scale set of intention prototypes and offers the ability to activate a sparse set of preferred intentions.

\begin{figure*}[t]
\centering
{\includegraphics[width=1.0\textwidth,height=0.5\textwidth]{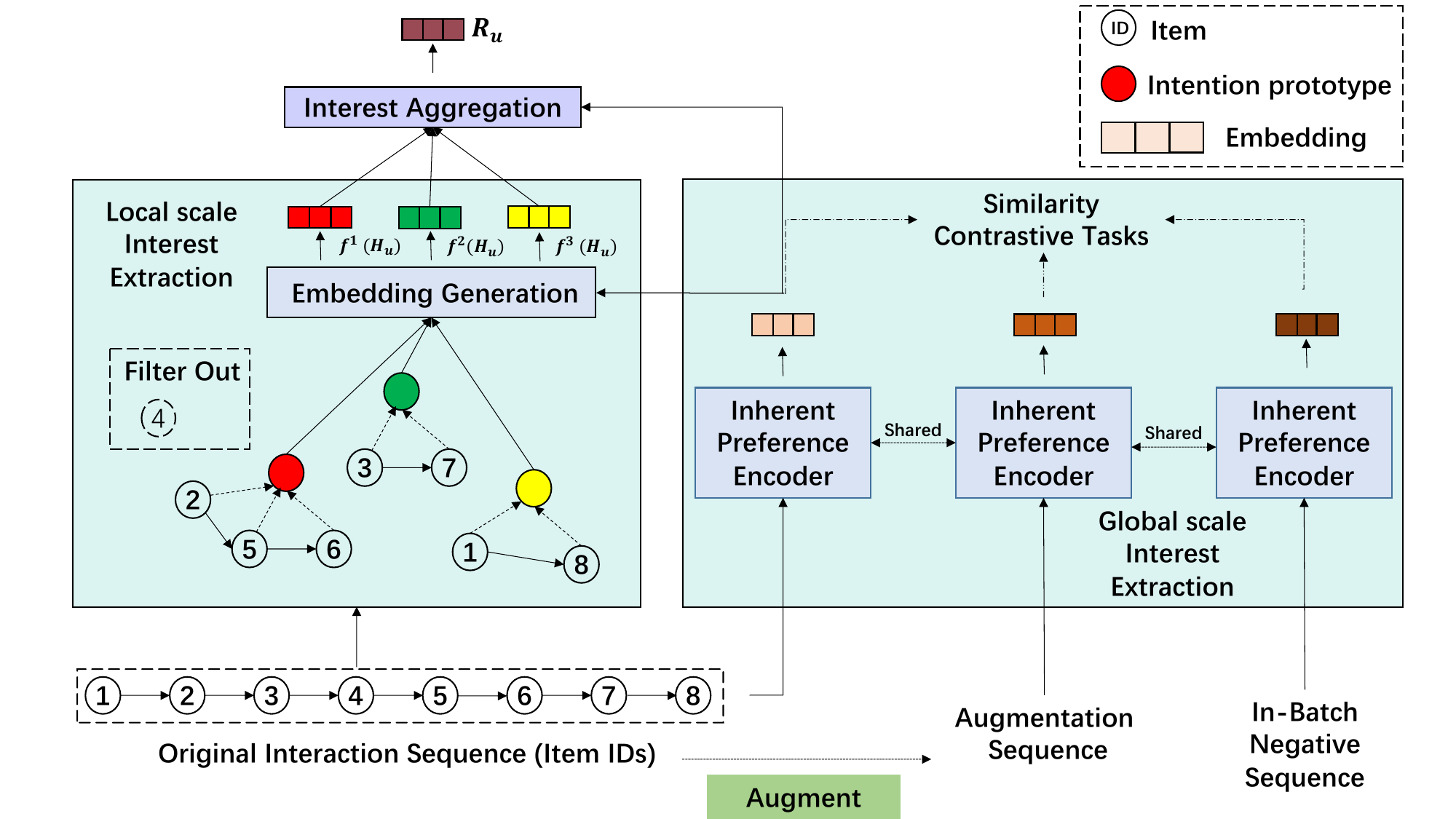}}
\caption{The architecture of DSIE (better viewed in color) consists of two scale interest extractions. The former captures the user's inherent preference at the global scale, and the latter extracts multiple interest embeddings at the local scale via intention prototype clustering, where Item $ID = 4$ is filtered out. Contrastive tasks on the similarity overcome the challenge of lacking explicit supervision for inherent preference modeling.}

\label{main-model} 
\end{figure*}

\section{THE PROPOSED METHOD}

In this section, we describe our DSIE in detail and explain the reason why we put forward the concept of dual-scale modeling. The overall framework is shown in Figure.\ref{main-model}, which includes: 1) Global scale Interest Extraction, which encodes the entire interaction sequence at the global scale to generate the inherent preference representation with similarity contrastive tasks; 2) Local scale Interest Extraction, which extracts multiple interest embeddings based on the corresponding sub-sequences discovered by intention prototype clustering at the local scale to represent different aspects of user's current interests separately; 3) Interest Aggregation, which adaptively integrates user's multiple interests according to the inherent preference to predict the user's current interest.

\subsection{Problem Formulation}
In a sequential recommendation task, we have a set of users $\mathcal{U}=\{u_{1},u_{2},...,u_{|\mathcal{U}|}\}$ and a set of items  $\mathcal{V}=\{v_{1},v_{2},...,v_{|\mathcal{V}|}\}$. Given a user $u \in \mathcal{U}$ and his/her historical interaction sequence $H_{u}=[v^{(u)}_{1},...,v_{i}^{(u)},...,v_{L}^{(u)}]$, where $v_{i}^{(u)}$ is the $i$-th item the user $u$ interacted with and $L$ is the length of the interaction sequence, the task is to predict the next item which the user $u$ would like to interact with.
, it can be expressed as: 

{\setlength{\abovedisplayskip}{0.05cm}
\setlength{\belowdisplayskip}{0.05cm}
\begin{equation}
    v_{\hat x} = \underset{x}{\operatorname{argmax}}   \mathcal{P}(v_{L+1}^{(u)} = v_x | H_{u})
\end{equation}}


However, due to constrained computing resource budgets and the low time-latency requirements, current industrial practice for recommender systems contains two stages: first retrieve top-${\rm N}$ relevant items from large-scale candidate items in the matching stage, and then rank them precisely in the ranking stage. Hence, the performance of the matching stage is an upper bound of the subsequent ranking stage and final recommendation accuracy. 
Here we take an effort in improving the effectiveness of matching models like the previous work  \cite{li2019multi,cen2020controllable,tan2021sparse}.

\subsection{Inherent Preference Modeling (Global Scale)}
We notice that each interest extraction in the recent multi-interests driven approaches works solely and is independent while oversimplifying the global correlation among the user's entire interacted items.
As the Fig.\ref{multi-interest} shows, the multi-interest extraction splits the original behavioral sequence into several sub-sequences, and each interest extraction is built on a sub-sequence. 
In other words, such a paradigm ignores the correlation of the user's intention at the lipstick and the dress in Fig.\ref{multi-interest}, which cannot understand the interaction sequence from an overall-level perspective to model the inherent preference and may fail
to jump out of the user's recent interacted item categories to explore the user's potential interests.


\subsubsection{Inherent Preference Encoder}
To address this issue, we capture the user's inherent preference from his/her entire user interaction sequence via the multi-layer ResNet \cite{he2016deep} and the self-attention \cite{vaswani2017attention, lin2022amif}.
The definition of $S$-layer ResNet $g(\cdot)$ is as follows:


{\setlength{\abovedisplayskip}{0.05cm}
\setlength{\belowdisplayskip}{0.05cm}
\begin{equation}
    \begin{aligned}
{h}_{1} &=\operatorname{ReLU}\left({\operatorname{Self-Atten}(E_u)} {W}_{r,1}+{b}_{r,1}\right)+{E_u} \\
{h}_{2} &=\operatorname{ReLU}\left(\operatorname{Self-Atten}({h}_{1}) {W}_{r,2}+{b}_{r,2}\right) +{h}_{1} \\
&  \cdots \\
{h}_{S} &=\operatorname{ReLU}\left(\operatorname{Self-Atten}({h}_{S-1}) {W}_{r,S}+{b}_{r,S}\right)+{h}_{S-1}
\end{aligned}
\label{tmw}
\end{equation}}

where ${E}_{u} = [e^{(u)}_{1},...,e_{i}^{(u)},...,e_{L}^{(u)}]$ 
denotes the initial item embedding matrix of 
$H_u$, ${W}_{r,*} \in \mathbb{R}^{d \times d}$, ${b}_{r,*} \in \mathbb{R}^{d}$
represent weight matrices, and bias terms, respectively. $d$ is the hidden dimension. The output is a sequence of hidden states $[{h}_{1}, {h}_{2}, ..., {h}_{S}]$, which captures potential semantic consistency under the entire interaction sequence from low- to high-orders.
The final learned inherent preference representation is a weighted aggregation of the entire historical interactions, formulated as follows:

{\setlength{\abovedisplayskip}{0.05cm}
\setlength{\belowdisplayskip}{0.05cm}
\begin{equation}
\begin{aligned}
{A}&=\operatorname{softmax}\left(\tanh \left([{h}_{1};  ...; {h}_{S}] {W}_{g,1} \right) {W}_{g,2}\right) \\
g(H_u) &= AE_u
\label{order}
\end{aligned}
\end{equation}}

where $[;]$ means the concatenation operation, ${W}_{g,1} \in \mathbb{R}^{Sd \times d}$ and ${W}_{g,2} \in \mathbb{R}^{d \times L}$ represent weight matrices.

\subsubsection{Contrastive Learning Task}
Since the public recommendation data just contains user's feedback (e.g. purchases), we could only get the recommendation results, but not the causes behind them. So it is difficult to guide the encoder to distinguish whether a particular user-item interaction is dominantly driven by the user's inherent preference or not.
For instance, the young girl in Fig.\ref{multi-interest} may purchase a nice phone case based on her inherent preference, while another one purchases it because many other users have purchased it.

Here we propose to use the self-supervised learning to overcome the challenge of lacking labeled cause-specific data. Specifically, inherent preference in the user's mind can be regarded as stable and insensitive to the order of his/her historical interactions. Hence we create the augmentation sequences based on the original sequence via item shuffling, formulated as $H_u \to H_u^{S}$. The pairwise contrastive task aims to maximize the similarity between $g(H_u)$ and $g( H_u^{S})$, while minimizing the similarity between the inherent preference representations $g(H_u)$ and $g(H_{u^{N}})$ in a contrastive manner~\cite{lin2022detect}, where $u^{N}$ denotes the in-batch negative sampling user. Formally, we implement the pairwise loss function based on Bayesian Personalized Ranking (BPR) \cite{rendle2012bpr} to accomplish contrastive learning, and use the inner product to measure embedding similarity, which is computed as follows:

{\setlength{\abovedisplayskip}{0.05cm}
\setlength{\belowdisplayskip}{0.05cm}
\begin{equation}
\mathcal{L}_{\mathrm{CL}}=\mathrm{BPR}(\langle g(H_u), g( H_u^{S})\rangle, \langle g(H_u), g(H_{u^{N}})\rangle)
\label{clloss}
\end{equation}}
where $\langle \cdot, \cdot \rangle$ denotes the inner product of two embeddings.






\subsection{Multi-Interest Extraction (Local Scale)}
As aforementioned, a single embedding vector is insufficient to express the user's diverse interests in practice. Therefore, this module is going to restructure the original interaction sequence $H_{u}$ into multiple sub-sequences, and then extract the user's diverse interests in a fine granularity from them at the local scale.



\subsubsection{Intention Clustering} 
 Following the idea of intention disentanglement as empirically proved in \cite{tan2021sparse, ma2019disentangled}, our multi-interest extraction module relies on a set of intention prototypes $C \in \mathbb{R}^{K \times d} $ identified via clustering, where $K$ is the number of intention prototypes.
 Towards that, we can estimate each item in the user's interaction sequence related to the intention prototypes according to their distances:

 

{\setlength{\abovedisplayskip}{0.05cm}
\setlength{\belowdisplayskip}{0.05cm}
\begin{equation}
P_{k \mid i}=\frac{\exp \left(\text { LN }_{1}\left({e}_i^{(u)} {W}_{c,1}\right) \cdot \text { LN }_{2}\left({C}_{k}\right)\right)}{\sum_{k^{\prime}=1}^{K} \exp \left(\text { LN }_{1}\left({e}_i^{(u)} {W}_{c,1}\right) \cdot \text { LN }_{2}\left({C}_{k^{\prime}}\right)\right)}
\label{k2i}
\end{equation}}

where $i = 1,2,...,L$ and $k = 1,2,..., K$, $P_{k \mid i}$ measures how likely the primary intention of the item at position $i$ is related to the $k$-th intention prototype, ${e}_i^{(u)}$ is the embedding of ${v}_i^{(u)}$ and ${C}_{k}$ is the embedding of the $k$-th intention prototype. 
${W}_{c,1} \in \mathbb{R}^{d \times d}$ is the trainable weight matrix. 
$\text { LN }_{j}(\cdot)$ represents a normalization layer and the subscript $j$ is used for identification the different layers. Given Equ.\ref{k2i}, those noisy behaviors in $H_{u}$ which are inconsistent with the user's real interests will be filtered out (i.e., $P_{k \mid i} \to 0$, for each $k$). 



\subsubsection{Position Weighting} In addition to the intention clustering weight $P_{k \mid i}$ calculated from the relevance perspective, we also consider another weight $P_{i}$ to discriminate the significance of items in different positions for estimating the user's current interest. The idea is motivated by the exact position of each interaction in the sequence playing a vital role in capturing the user's current interest. For instance, \cite{tan2021sparse} found that users tend to consume items that are similar to their recent consumption according to the statistics on the Taobao dataset \cite{zhu2018learning}.
Formally, we leverage another soft-attentive layer to measure the position importance, formulated as follows:





{\setlength{\abovedisplayskip}{0.05cm}
\setlength{\belowdisplayskip}{0.05cm}
\begin{equation}
    \begin{aligned}
\begin{gathered}
P_i=\operatorname{softmax}\left(\operatorname{ReLU} \left({E}_{u} {W}_{p,1} + b_{p,1}\right) {W}_{p,2} + b_{p,2}\right)^{\top}
\end{gathered}
\end{aligned}
\end{equation}}

where ${W}_{p,1} \in \mathbb{R}^{d \times 4d} $, ${W}_{p,2} \in \mathbb{R}^{4d \times K} $, $b_{p,1} \in \mathbb{R}^{ 4d}$ and $b_{p,2} \in \mathbb{R}^{ d}$ are parameters.



\subsubsection{Inherent Preference-Guided Attention Network}
In most cases, the user's interactions are noisy, including the invalid interactions derived from accidental clicks rather than user's real interests.
Therefore, it is of significance to pay different attention to these interacted items in order to accurately estimate the user's real interests.
It is intuitive that the user's inherent preference could help our framework focus on the items reflecting the real interests.
Towards that, given the initial item embedding matrix $E_u$ and the user's inherent preference representation $g(H_u)$, we leverage a guided attention network~\cite{lin2021rumor} to calculate the relevance of each interacted item with respect to the inherent preference as follows:

\begin{equation}
a_i=\operatorname{sigmoid}\left({W}_{pr,2}^{\top}\operatorname{tanh}\left({W}_{pr,1}^{\top}\left[{e}_i^{(u)}; g(H_u)\right]\right)\right)
\end{equation}
where ${W}_{pr,2} \in \mathbb{R}^{2d \times d} $, ${W}_{pr,1} \in \mathbb{R}^{d} $ are weight matrices.

\subsubsection{Multi-Interest Embedding Generation} We can generate multiple interest embeddings from the user's behavior sequence $H_u$ according to $P_{k \mid i}$, $P_{i} $, and $a_i$.
Formally, the $k$-th output of our multi-interest extraction is computed as follows:

{\setlength{\abovedisplayskip}{0.05cm}
\setlength{\belowdisplayskip}{0.05cm}
\begin{equation}
f^{k}\left({H}_{u}\right)=\text { LN }_{3}\left(\sum_{i=1}^{L} P_{k \mid i} \cdot P_{i } \cdot {a}_{i} \cdot {e}_{i}^{(u)} + b_{m,k}\right)
\label{multi-int-calc}
\end{equation}}

where $f^{k}(\cdot)$ is the $k$-th extraction function, and $b_{m,k} \in \mathbb{R}^d$ is the bias term.


 
\subsection{Interest Aggregation Module}


On top of the multi-interest extraction module, we obtain multiple interest embeddings to express the diverse interests of a user. However, the multi-interest extraction splits the original behavioral sequence into $K$ sub-sequences $\left\{H_{u}^{k}\right\}_{k=1}^{K}$ according to $P_{k \mid i}$, and each interest extraction is built on a sub-sequence (i.e., local scale) $H_{u}^{k}$ related to an intention prototype $C_k$.
Intuitively, multi-interests can be regarded as the specific interests derived from inherent preference when the user interacts with the specific items.
In sight of this, we develop an interest aggregation module to adaptively integrate the multi-interests according to the inherent preference to predict the user's current interest. Formally, the final fused interest $R_u $ is formulated as :

{\setlength{\abovedisplayskip}{0.05cm}
\setlength{\belowdisplayskip}{0.05cm}
\begin{equation}
\begin{aligned}
\alpha_{k}^{u}&=\frac{\exp \left(\left(g(H_u)\right)^{\top} f^{k}\left({H}_{u}\right) / \tau\right)}{\sum_{k^{\prime}=1}^{K} \exp \left(\left(g(H_u)\right)^{\top}
f^{k^{\prime}}\left({H}_{u}\right)) / \tau\right)} \\
R_u&=\sum_{k=1}^{K} \alpha_{k}^{u} \cdot f^{k}\left({H}_{u}\right)
\label{finalinterest}
\end{aligned}
\end{equation}}



where 
 $\alpha^{u} = [\alpha_{1}^{u}, \alpha_{2}^{u}, ..., \alpha_{K}^{u}]^{\top} \in R^{K}$ is the attention vector for diverse interests. If the specific interest is more correlated with the inherent preference, its weight in the aggregation will be more significant, and vice versa. 
$\tau$ is a hyper-parameter that controls the smoothness of the output distribution, and we set $\tau = 0.1 $. 

\subsection{Model Optimization}
After the dual-scale interest extraction framework, we obtain the representation of the user's current interest, which can be exploited to retrieve top-${\rm N}$  relevant items during the inference. During the training, a softmax layer is employed to produce the distribution of matching score between the user and a specific item:

{\setlength{\abovedisplayskip}{0.05cm}
\setlength{\belowdisplayskip}{0.05cm}
\begin{equation}
P(v_x) = \log \frac{\exp (R_{u}^{\top} e_x)}{\left.\sum_{j \in\{1,2, \cdots, |\mathcal{V}|\}} \exp (R_{u}^{\top} e_j)\right)}
\label{softmax}
\end{equation}}

where $e_x$ is the embedding of the specific item $v_x$ shared with the embedding layer for reducing model size. We leverage a sampled softmax technique \cite{ covington2016deep} here as an alternative to reduce the computational complexity of the sum operation of the denominator in Equ.\ref{softmax}. The objective function of our model is to minimize the negative log-likelihood as follows:
\begin{equation}
    \mathcal{L}= \frac{1}{|\mathcal{U}|}\sum_{u \in \mathcal{U}} \sum_{t=2}^{L}-\log P\left(v_{x} = v_t^{(u)} \mid H_{u,t}\right)
\end{equation}
where $H_{u,t} = [v_1^{(u)}, v_2^{(u)}, ..., v_{t-1}^{(u)}]$ denotes that each training sample uses the first $t-1$ interactions as prior truncated sequence to predict $v_t^{(u)}$ following the common practice in \cite{kang2018self, li2019multi}.

Besides, we introduce a regularization method to impose the learned K interest embeddings $\left\{f^{k}(H_u)\right\}_{k=1}^{K}$ 
to preserve sufficiently different information. Following \cite{cogswell2015reducing}, we propose to impose the intention prototypes $C$ orthogonally, i,e., $T=\frac{1}{K} (C - \bar{C}) (C - \bar{C})^{\top}$, $\mathcal{L}_{aux}=\frac{1}{2}\left(\|{T}\|_{F}^{2}-\|\operatorname{diag}({T})\|_{2}^{2}\right)$, where $T$ is the covariance matrix of intention prototypes $C$, $\|\cdot\|_F$ denotes the Frobenius norm, and the $\operatorname{diag}(\cdot)$ extracts the main diagonal of a matrix into a vector.
We minimize the total loss $\mathcal{L} + \alpha\mathcal{L}_{aux} + \beta \mathcal{L}_{\mathrm{CL}}$ with two hyper-parameters $\alpha$ and $\beta$ to balance objectives.

\section{EXPERIMENTS}

In this section, we introduce our experimental settings and report our empirical results on three benchmark datasets. Our experiments are designed to answer the following research questions. RQ1: Can our proposed DSIE outperform state-of-the-art baselines for sequential recommendation tasks?  RQ2:How sensitive are the hyper-parameter settings, including the temperature parameter $\tau$ in the interest aggregation module, the number of multiple interest embeddings $K$, and the layer number of multi-layer ResNet $S$? RQ3: Does the proposed method also help in improving the performance of recommending the novel items?






\begin{table}
\caption{Dataset statistics (after preprocessing).}
\centering
\resizebox{0.48\textwidth}{!}{
\begin{tabular}{cccccc}
\hline
Dataset & \#Users & \#Items & \#Interaction & \#Avg.length \\ \hline \hline
Video Games & 24303& 10672&0.23m & 9.537 \\
   Movie\&TV  & 123960 & 50052&1.70m & 13.694 \\
   {Kindle Store} & 139785& 98824&2.22m & 15.901\\\hline
\end{tabular}}
\label{datasets}
\end{table}

\subsection{Setup}
\subsubsection{Datasets}
We evaluate the proposed model on three real-world representative datasets which vary significantly in domains from the Amazon dataset \cite{mcauley2015image}, including "Games", "Movies" and  "Kindle" categories. The dataset also includes rich information that might be useful in our recommendation novelty analysis, like user's reviews, item categories, etc. 
To reduce data sparsity, we filtered out all the users and items that have less than three instances of feedback. Table 1 lists the statistics of three datasets.

\subsubsection{Task Settings}
To evaluate the sequential recommendation models under strong globalization, We split all users into training/validation/test sets by the proportion of 8:1:1, which has been widely used in \cite{ma2019learning}. We train models using the entire interaction sequences of training users. To evaluate, for each user in validation/test sets, we hold out the last 20\% of the user's interactions as the test data to compute metrics and utilize the remaining 80\% items for inferring user representation from trained models. To evaluate the performance of all the models, we report Recall@N, NDCG@N, and HR@N with ${\rm N} = 50$.
\begin{table*}
\caption{Performance comparison of different methods on public datasets.
Bold scores are the best in each row, while underlined scores are the second best. Improvements are shown in the last column. $*$ indicates the significant improvement over all baselines with $p$-value $<$ 0.01. All the numbers in the table are percentage numbers with '\%' omitted.}

\centering
\resizebox{1.0\textwidth}{!}{
\begin{tabular}{ccccccccccc}
\hline
Dataset & Metric & MostPopular & YoutubeDNN & GRU4Rec  & MIND & ComiRec-SA & ComiRec-DR & SINE& DSIE & Improv. \\ \hline \hline
 \multirow{3}{*}{\text{Video Games}} 
 
            
          

           & Recall@50 & 7.260  & 15.670 & 13.932 & 18.062 &  18.667 &  18.332  & \underline{21.194} & $\mathbf {23.512}^{*}$  & +10.937  \\
           
          ~ & NDCG@50 & 4.016 & 9.004  & 8.219 & 10.074 & 9.511 & 9.911 & \underline{10.117}   & $\mathbf {11.830}^{*}$& +16.932  \\
          
          ~ & HR@50 & 13.821 & 26.820 &  23.776 & 30.440 & 30.522 & 30.734 & \underline{34.513} & $\mathbf {37.309}^{*}$   & +8.101  \\

     \multirow{3}{*}{\text{Movies\&TV}} 
     
            
          

           & Recall@50 & 5.536  & 10.564 & 9.449 & 12.040 &  11.748 &  11.942 & \underline{13.318}  & $\mathbf {14.852}^{*}$ & +11.518  \\
           
          ~ & NDCG@50 & 3.637 & 6.850  &5.955 & 7.333 & 6.327 & 6.864 & \underline{7.683} & $\mathbf {8.790}^{*}$  & +14.408  \\
          
          ~ & HR@50 & 11.560 & 20.313 & 18.135 & 22.249 & 21.055 & 22.039 & \underline{23.992}& $\mathbf {26.751}^{*}$  & +11.500   \\
            
     \multirow{3}{*}{\text{Kindle Store}}
     
            
          

          & Recall@50 & 1.821  & 10.564 & 9.449 & 16.731 &  16.192 &  16.492 & \underline{17.372}   & $\mathbf {18.505}^{*}$  & +6.522  \\
           
          ~ & NDCG@50 & 3.094 & 6.850  &5.955 & 9.726 & 9.811 & 9.957 & \underline{11.188}  & $\mathbf {12.266}^{*}$& +9.635  \\
          
          ~ & HR@50 & 4.407 & 20.313 & 18.135 & 31.805 & 30.657 & 30.923 & \underline{31.862}   & $\mathbf {33.715}^{*}$   & +5.816  \\ \hline
        
\end{tabular}}
\label{performance}
\end{table*}

\subsubsection{Baselines}
To verify the effectiveness of DSIE, we compare it with the following representative baselines: 1) MostPopular: A traditional recommendation method that recommends the most popular items to users. 2) YoutubeDNN \cite{covington2016deep}: It is one of the most successful deep learning models for industrial recommender systems. 3) GRU4Rec \cite{hidasi2015session}: It is the first work that introduces recurrent neural networks to model user behaviors. 4) MIND \cite{li2019multi}: It designs a multi-interest extractor layer based on the capsule routing mechanism, which is applicable for clustering past behaviors and extracting diverse interests. 5) ComiRec-SA  \cite{cen2020controllable}: It leverages a self-attentive method for multi-interest extraction. 6) ComiRec-DR \cite{cen2020controllable}: It improves the original dynamic routing method used by capsule network, compared with MIND.
 7) SINE \cite{tan2021sparse}: It is a state-of-the-art multi-interest framework based on latent conceptual prototypes.

\subsubsection{Implementation Details}
For all baselines, we use the code and the default settings of hyper-parameters according to the respective papers and fine-tune them using the validation set and terminated training if validation performance didn't improve within 20 epochs to be fair. We implement our framework DSIE with TensorFlow. We set the hidden dimension $d = 128$, the maximum sequence length $L = 20$, the loss weights $\alpha = 0.1$, $\beta = 0.4$ for all datasets, and the number of samples for sampled softmax loss is set to 10. 
The layer number of ResNet $S$ is 2 for Kindle Store and 3 for other datasets. The number of intention prototypes $K$ is 3 for Video Games, 5 for Movies\&TV, and 4 for Kindle Store.
The parameters are updated by the Adam \cite{kingma2014adam} optimizer with the learning rate initialized as 0.001. All the models are trained from scratch without any pre-training on a single GTX-1080 Ti GPU with a batch size of 128.


\subsection{Overall Performance Comparison (RQ1)}
\subsubsection{Main Results}
Table \ref{performance} shows the recommendation performance of all methods on three public datasets. For all datasets, multi-interests based recommenders (e.g., MIND, ComiRec-SA, and SINE) have obviously better performance than single interest based recommenders (e.g., YoutubeDNN, and GRU4Rec) because these methods have a stronger express ability to capture the user's diverse interests which has proved to be an effective way of estimating their current interest and boosting sequential recommendation accuracy. Among the baseline methods, SINE \cite{tan2021sparse} has state-of-the-art performance compared with other baselines. 
This makes sense since SINE adaptively activates multiple intentions from a large pool of conceptual prototypes to generate multiple interest embeddings for a user explicitly. 

Clearly, DSIE improves over the best baseline methods on all datasets with respect to all metrics. Possible reasons are that our model takes advantage of intention prototypes to extract the multi-interest explicitly. Compared with the single interest based recommenders (e.g., YoutubeDNN, and GRU4Rec) and the implicit multi-interest based recommenders (e.g., MIND), our method can effectively filter out noisy behaviors via intention clustering and be more focused on the user's real interest modeling. 
Besides, DSIE captures the user's inherent preference by understanding the global correlation of their entire interaction sequence with self-supervised learning and adopts an adaptive aggregation module to aggregate the multi-interests according to the inherent preference, which achieves a more precise representation of user's current interest.

\begin{table}
\caption{Ablation study of DSIE.}
\centering
\resizebox{0.49\textwidth}{!}{
\begin{tabular}{cccccc}
\hline
 Dataset & Metric   &  DSIE w/o GS & DSIE w/o CL & DSIE \\ \hline \hline
    \multirow{3}{*}{\text{Video Games}} & Recall@50   & 21.012& $ {22.075}$ & $ {23.512}$   \\
            
           ~ & NDCG@50  & 9.423 & {10.751}  & $ {11.830}$ \\
           
           ~ & HR@50  & 34.634 & $ {35.376}$  & $ {37.309}$    \\

    \multirow{3}{*}{\text{Movies\&TV}} & Recall@50   & 13.189 & $ {13.611}$ & $ {14.852}$ \\
             ~ & NDCG@50  & 7.539 & $ {7.966}$ & $ {8.790}$ \\
           ~ &  HR@50 & 23.974 & $ {24.823}$ & $ {26.751}$  \\
           
    \multirow{3}{*}{\text{Kindle Store}} & Recall@50   &17.178 & $ {17.616}$ & $ {18.505}^{*}$  \\
             ~ & NDCG@50  & 10.901 & $ {11.652}$ & $ {12.266}$\\
           ~ & HR@50 & 31.982 & $ {32.542}$ & $ {33.715}$    \\
          \hline
\end{tabular}}
\label{ablation}
\end{table}

\begin{figure*}[htbp]
\centering
\subfigure[temperature parameter $\tau$.]{
\begin{minipage}[t]{0.33\linewidth}
\centering
\includegraphics[width=2.3in]{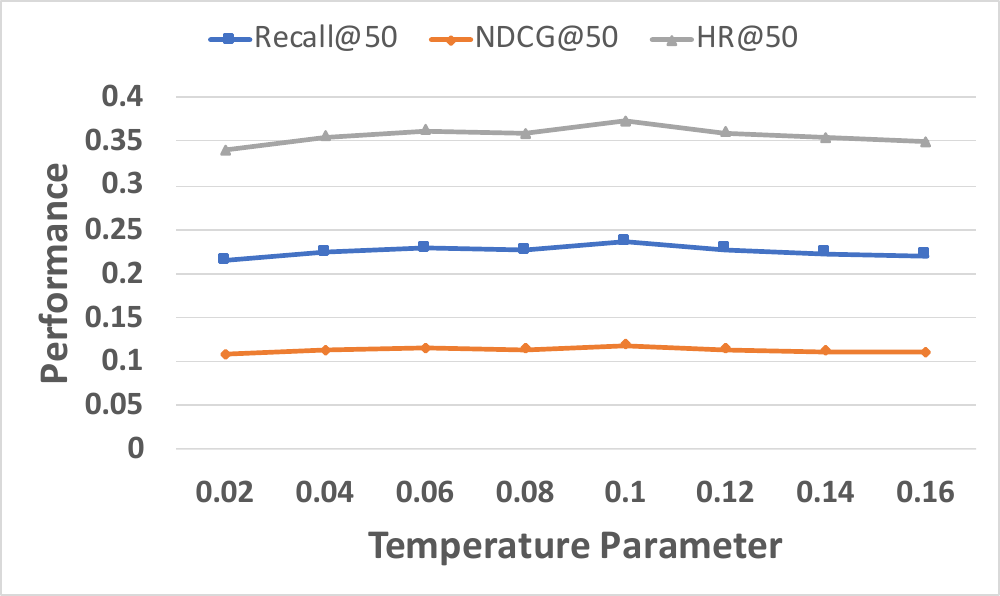}
\end{minipage}%
}%
\subfigure[ResNet layer num $S$.]{
\begin{minipage}[t]{0.33\linewidth}
\centering
\includegraphics[width=2.3in]{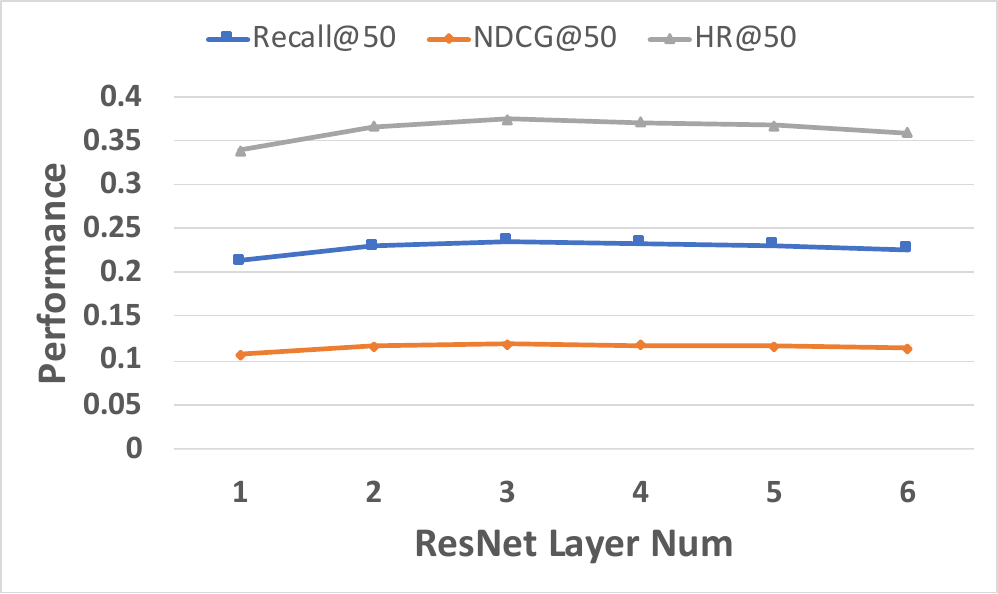}
\end{minipage}%
}%
\subfigure[interest embeddings num $K$.]{
\begin{minipage}[t]{0.33\linewidth}
\centering
\includegraphics[width=2.3in]{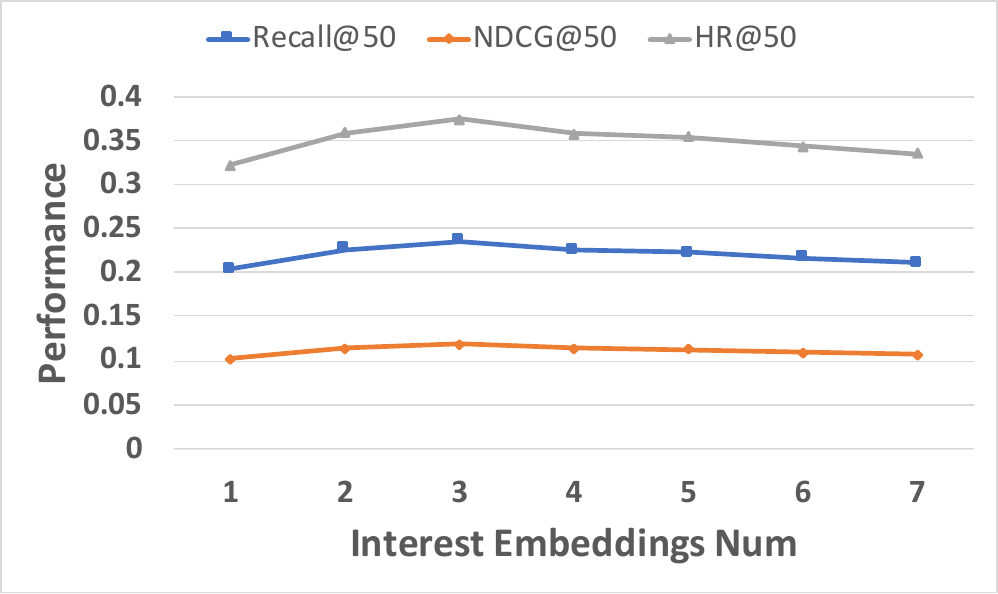}
\end{minipage}
}%
\centering
\caption{Effect of several hyper-parameters on ranking performance (Metrics@50).}
\label{param_sensi}
\end{figure*}

\subsubsection{Ablation Study}
We introduce the ablation study to compare the performance of DSIE with two variants (DSIE w/o GS and DSIE w/o CL) to validate the effectiveness of the inherent preference modeling (i.e., global-scale) and the contrastive learning loss $\mathcal{L}_{\mathrm{CL}}$ in Table \ref{ablation}. Specifically, DSIE w/o GS leverages the fine-grained multi-interest extraction only and retrieves $K \times N $ items based on $K$ output interest embeddings first and then ranks them to get the final top-${\rm N}$ items as the recommendation results like MIND.
Obviously, DSIE significantly outperforms the variant (DSIE w/o CL) in three datasets to verify that our contrastive tasks on the similarity between the learned representations of the original sequence, the augmentation sequences, and the in-batch negative sequences help overcome the challenge of lacking labeled cause-specific data as explicit supervision and achieve stronger performance than existing unsupervised methods (e.g., MIND, ComiRec-SA, and SINE). Moreover, the improvement of the variant (DSIE w/o CL) over DSIE w/o GS validates that thanks to the awareness of the user's inherent preferences, the variant (DSIE w/o CL) is capable to capture the user's current interest more precisely.

\subsection{Hyper-parameter Sensitivity (RQ2)}
In this section, we conduct three experiments on Video Games dataset to study the influence of the hyper-parameters within the multi-interest extractor layer, the multi-layer ResNet, and the interest aggregation module. Here we summarize the performance of our model in terms of metrics@50 in Figure \ref{param_sensi}. 

\subsubsection{Influence of temperature parameter $\tau$ in interest aggregation module} As mentioned before, the temperature parameter $\tau$ controls the balance of the correlation between each interest and the inherent preference. We compare the performance of DSIE as $\tau$ varies from 0.02 to 0.16 and show the results in the left part of Figure \ref{param_sensi}. Clearly, the performance of $\tau = 0.02$ is much worse than the others. The reason is that, when taking $\tau \to 0$, the most relevant interest dominates the aggregation, and the multi-interest extraction degenerates the single interest based recommender (e.g., YoutubeDNN and GRU4Rec, etc).  It also shows that performance gets better as $\tau$ increases, and obtains the best performance when $\tau = 0.1$. When $\tau$ keeps increasing, the performance declines. We attribute this to when taking $\tau \to +\infty$, each interest has the same attention thus the aggregation module equals the average of interests with no reference to the inherent preference. Thus, DSIE achieves its best performance when the temperature parameter is chosen properly $\tau = 0.1$.

\subsubsection{Influence of the layer number of multi-layer ResNet $S$} The middle part of Figure \ref{param_sensi} illustrates that DSIE has stable performance under different layer numbers of multi-layer ResNet $S$. 
Not surprisingly, results are inferior with one ResNet layer, and stacking the ResNet layer can boost performances, which verifies that deep architecture is helpful to learn more complex features. 
The variant with two layers performs reasonably well, and using three layers achieves its best performance.
However, larger $S$ means more parameters need to be trained and the decline is largely due to overfitting.

\subsubsection{Influence of the number of multiple interest embeddings $K$} As aforementioned, the number of multiple interest embeddings $K$ is a key factor in model performance, which directly affects the model's expression capacity. 
We vary the number of multiple interest embeddings from 1 to 7 and report the results in the right part of Figure \ref{param_sensi} as. From the results, we find that $K$ should not be too small or it is not enough to capture the user's diverse interests. Meanwhile, it should not be too large, otherwise, it would be hard to converge since it is inconsistent with the real-world recommendation situation. 
Specifically, as $K$ increases from 1 to 3, DSIE starts to achieve a greater performance, which emphasizes the importance of leveraging multiple embeddings to increase model expression capacity. However, when $ K > 3$, the performance brought by DSIE starts to fall. These results verify what we claimed above.

\subsection{Recommendation Novelty Analysis (RQ3)}

            
           

           

Here we carry out the recommendation novelty analysis to highlight the contribution of the user's inherent preference modeling. To validate the recommendation novelty, a common way in practice is to filter those interacted items in the prediction sets \cite{devooght2016collaborative,liang2016factorization,zhu2018learning}, i.e., only those novel items could be ultimately recommended. 
Thus, recommendation novelty depends on comparing accuracy in a complete novel prediction set. 
In our experiments, we filter items of the same item category as the user's historical behaviors in all methods' prediction sets, based on the metadata of the Amazon dataset \cite{mcauley2015image}. 
For example, the category of \emph{"Darkside Blues VHS"} is \emph{"Animation"}, so we filter all items whose item category is \emph{"Animation"} in the prediction set.
It's worth mentioning that the prediction set size will be complemented to the required number ${\rm N}$ if its size is smaller than ${\rm N}$ after filtering. We adopt the same way of generating the prediction sets for all methods to guarantee fair comparison. 

Table \ref{performance-filter} demonstrates that DSIE outperforms the existing multi-interest driven methods \cite{li2019multi,cen2020controllable,tan2021sparse} in a complete novel prediction set. \footnote{Our empirical results in Table \ref{performance} and the original paper both show that the performance of ComiRec-SA and ComiRec-DR is comparable. Here we only report the ComiRec-SA's results for brevity.}
This indicates that compared with the baselines usually produce homogeneous and no-diverse recommendation predictions, our model recommends highly related but novel items. It is reasonable since being aware of the user's inherent preferences, DSIE is able to jump out of the user's 
recent interacted item categories to explore the user's potential interests, so as to produce diverse recommendation lists and improve the novelty.

 



\begin{table}
\caption{Recommendation novelty comparison of different methods.}
\centering
\resizebox{0.48\textwidth}{!}{
\begin{tabular}{cccccc}
\hline
 Dataset & Metric  & MIND  &  ComiRec-SA & SINE & DSIE \\ \hline \hline
    \multirow{3}{*}{\text{Video Games}} & Recall@50   & 4.529& 4.487& {6.581}  & $  {6.786}$   \\
            
           ~ & NDCG@50  & 2.371& 2.332 & {2.654}  & $  {2.817}$ \\
           
           ~ & HR@50  & 10.778 & 10.107 & {12.834}  & $  {13.040}$ \\

    \multirow{3}{*}{\text{Movies\&TV}} & Recall@50   & 5.354& 5.289 & {5.954}  & $  {6.245}$  \\
             ~ & NDCG@50  & 2.670& 2.539 & { 2.991}  & $  {3.191}$ \\
           ~ &  HR@50 & 11.302 & 10.974 & {12.932}  & $  {13.553}$  \\
           
    \multirow{3}{*}{\text{Kindle Store}} & Recall@50   &3.233 &3.078 & {3.637}  & $  {3.805}$  \\
             ~ & NDCG@50  & 1.953 & 1.801 & {1.845}  & $  {2.030}$ \\
           ~ & HR@50 & 6.195 & 6.382 & {6.713}  & $  {6.826}$  \\
          \hline
\end{tabular}}
\label{performance-filter}
\end{table}

\section{CONCLUSION}
In this work, we proposed a novel dual-scale interest extraction framework (DSIE) for the sequential recommendation. 
DSIE exploits an inherent preference encoder with self-supervised learning to capture the global correlation of the entire interaction sequence (i.e., global scale), and a multi-interest extraction layer to model the user's diverse interests based on the sub-sequences (i.e., local scale) discovered via the intention prototype clustering. 
Besides, an adaptive interest aggregation is introduced to integrate the multi interest according to the inherent preference to generate the user's current interest.
Experimental results on three public datasets show that our model outperforms state-of-the-art baselines in terms of recommendation accuracy and novelty.

\bibliography{ecai}
\end{document}